\documentclass[12pt]{article}

\usepackage{latexsym}
\usepackage{amsmath,amsfonts}
\usepackage{times}
\allowdisplaybreaks[4]

\catcode`\@=11

\newcount\hour
\newcount\minute
\newtoks\amorpm \hour=\time\divide\hour by 60\minute
=\time{\multiply\hour by 60 \global\advance\minute by-\hour}
\edef\standardtime{{\ifnum\hour<12 \global\amorpm={am}%
        \else\global\amorpm={pm}\advance\hour by-12 \fi
        \ifnum\hour=0 \hour=12 \fi
        \number\hour:\ifnum\minute<10
        0\fi\number\minute\the\amorpm}}
\edef\militarytime{\number\hour:\ifnum\minute<10 0\fi\number\minute}

\def\draftlabel#1{{\@bsphack\if@filesw {\let\thepage\relax
   \xdef\@gtempa{\write\@auxout{\string
      \newlabel{#1}{{\@currentlabel}{\thepage}}}}}\@gtempa
   \if@nobreak \ifvmode\nobreak\fi\fi\fi\@esphack}
        \gdef\@eqnlabel{#1}}
\def\@eqnlabel{}
\def\@vacuum{}
\def\marginnote#1{}
\def\draftmarginnote#1{\marginpar{\raggedright\scriptsize\tt#1}}
\def\draft{
        \pagestyle{plain}
        \overfullrule=2pt
        \oddsidemargin -.5truein
        \def\@oddhead{\sl \phantom{\today\quad\militarytime} \hfil
        \smash{\Large\sl DRAFT} \hfil \today\quad\militarytime}
        \let\@evenhead\@oddhead
        \let\label=\draftlabel
        \let\marginnote=\draftmarginnote
        \def\ps@empty{\let\@mkboth\@gobbletwo
        \def\@oddfoot{\hfil \smash{\Large\sl DRAFT} \hfil}
        \let\@evenfoot\@oddhead}
        \def\@eqnnum{(\theequation)\rlap{\kern\marginparsep\tt\@eqnlabel}%
        \global\let\@eqnlabel\@vacuum}  }

\newcommand{\rf}[1]{(\ref{#1})}
\renewcommand{\theequation}{\thesection.\arabic{equation}}
\renewcommand{\thefootnote}{\fnsymbol{footnote}}
\newcommand{\newsection}{   
\setcounter{equation}{0}\section}

\def\appendix#1{\addtocounter{section}{1}\setcounter{equation}{0}
\renewcommand{\thesection}{\Alph{section}}
\section*{Appendix \thesection\protect\indent \parbox[t]{11.15cm}{#1}}
\addcontentsline{toc}{section}{Appendix \thesection\ \ \ #1}}

\def\be{\begin{equation}}
\def\ee{\end{equation}}
\def\beq{\begin{eqnarray}}
\def\eeq{\end{eqnarray}}

\def\parline{\,\partial\kern -0.55em /\,\,}

\def\half{{\frac{1}{2}}}

\def\CC{{\cal C}}

\def\PP{{\cal P}}

\def\abf{{\bf a}}
\def\bbf{{\bf b}}
\def\cbf{{\bf c}}
\def\dbf{{\bf d}}
\def\ebf{{\bf e}}
\def\fbf{{\bf f}}

\def\pbf{{\bf p}}

\def\adbf{{\bf ad}}

\def\ibf{{\bf i}}
\def\iibf{{\bf ii}}
\def\iiibf{{\bf iii}}
\def\ivbf{{\bf iv}}
\def\vbf{{\bf v}}
\def\vibf{{\bf vi}}

\def\qqq#1#2{\begin{array}{l}
#1
\\[2pt]
#2
\end{array} }

\def\Rsm{{\scriptscriptstyle R}}
\def\Lsm{{\scriptscriptstyle L}}

\def\rh{{\hat{r}}}

\def\rrr{r}

\newcommand{\Eo}{\mathbb{E}}

\newcommand{\No}{\mathbb{N}}

\newcommand{\Ro}{\mathbb{R}}

\newcommand{\Zo}{\mathbb{Z}}

\def\sm(A)dS{{\scriptscriptstyle (A)dS }}

\def\brm{{\rm b}}
\def\frm{{\rm f}}
\def\irm{{\rm i}}

\def\flatrm{{\rm flat}}

\def\maxrm{{\rm max}}
\def\minrm{{\rm min}}

\def\specsf{{\sf spec}}

\def\noinbf#1{\noindent {\bf #1}}
\newcommand{\mc}{\multicolumn}

\begin{document}


\begin{flushright}
FIAN-TD-2025-15  \hspace{1.7cm} \ \ \ \  \\
arXiv: 2507.05194 V2 [hep-th] \\
\end{flushright}

\vspace{1cm}

\begin{center}

{\Large \bf Light-cone vector superspace and continuous-spin field in AdS}

\vspace{2.5cm}

R.R. Metsaev%
\footnote{ E-mail: metsaev@lpi.ru
}

\vspace{1cm}

{\it Department of Theoretical Physics, P.N. Lebedev Physical
Institute, \\ Leninsky prospect 53,  Moscow 119991, Russia }

\vspace{3.5cm}

{\bf Abstract}

\end{center}

In the framework of light-cone gauge vector superspace, a continuous-spin field propagating in AdS space of dimension greater than or equal to four is studied. Use of such framework allows us to find a simple solution for spin operators entering our light-cone gauge Lagrangian formulation. Bosonic continuous-spin field is considered in AdS space of arbitrary dimensions, while the use of a helicity basis in four-dimensional AdS space allows us to consider bosonic and fermionic continuous-spin fields on an equal footing. Classification of continuous-spin fields is proposed.
Conjectures on the notion of masslessness for continuous-spin field are made. All unitary irreps of non-linear spin algebra for four-dimensional AdS are also obtained.

\vspace{2cm}

Keywords: continuous-spin field; spin-algebra, AdS space.

\newpage
\renewcommand{\thefootnote}{\arabic{footnote}}
\setcounter{footnote}{0}

\newsection{\large Introduction}

Lagrangian formulation of a bosonic continuous-spin field (CSF) in flat space developed in Refs.\cite{Schuster:2013pta} has attracted some interest in the literature (for fermionic CSF see Ref.\cite{Najafizadeh:2015uxa}). Reviews and list of references on earlier studies in the topic of CSF may be found in Refs.\cite{Brink:2002zx}. For CSF in (A)dS space, a Lagrangian gauge invariant formulation was obtained in Refs.\cite{Metsaev:2016lhs,Metsaev:2017ytk} (see also Refs.\cite{Zinoviev:2017rnj}), while a light-cone gauge formulation was discussed in Refs.\cite{Metsaev:2017myp}-\cite{Metsaev:2021zdg}.
Supersymmetric CSF was investigated in Refs.\cite{Buchbinder:2019kuh}, while BRST studies of CSF may be found in Refs.\cite{Bengtsson:2013vra}-\cite{Burdik:2020ror}. Interacting CSF was studied in Refs.\cite{Metsaev:2017cuz}-\cite{Metsaev:2018moa}, while discussion of CSF in the framework of a worldline formalism may be found in Refs.\cite{Schuster:2023xqa} (see also Ref.\cite{Basile:2023vyg}). An arbitrary mixed-symmetry CSF was considered in Refs.\cite{Metsaev:2021zdg,Alkalaev:2017hvj}, while an unfolded formulations of CSF was investigated in Refs.\cite{Ponomarev:2010st}. Discussion of other various interesting aspects of CSF may be found in Refs.\cite{Schuster:2013pxj}-\cite{Takata:2024srr}.

In Ref.\cite{Metsaev:2025qkr}, we developed the light-cone gauge vector superspace formulation of CSF in flat space and applied such formulation for a study of interacting CSF. We found that the use of the light-cone gauge vector superspace considerably simplifies interaction vertices of CSF as compared to the ones obtained by using the oscillator formulation in Refs.\cite{Metsaev:2017cuz,Metsaev:2018moa}. This motivates us to develop the light-cone gauge vector superspace formulation of CSF in AdS space. In this paper, firstly, we study a bosonic CSF in $AdS_{d+1}$, $d>3$, secondly, using a helicity basis, we investigate bosonic and fermionic CSFs in $AdS_4$, and, thirdly, we obtain all unitary irreps of non-linear spin algebra in $AdS_4$. Our research convinces us that the use of the light-cone vector superspace indeed considerably simplifies spin operators which play a crucial role in the light-cone gauge Lagrangian formulation of CSF. Also our results provide us a classification of CSF in $AdS_{d+1}$, $d\geq 3$.

\vspace{-0.2cm}
\newsection{ \large Bosonic continuous-spin field in $AdS_{d+1}$, $d>3$}\label{lc-action}

\noindent {\bf Field content}. To discuss a light-cone gauge bosonic CSF in $AdS_{d+1}$, $d>3$, we introduce the following set of fields:%
\footnote{ We use the Poincar\'e parametrization of AdS: $ds^2=R^2(-dx_0^2+dx_i^2+dx_{d-1}^2+dz^2)/z^2$ and light-cone coordinates $x^\pm$,  $x^\pm:=(x^{d-1} \pm x^0)/\sqrt{2}$, where
$x^+$ stands for the light-cone time. For the coordinates and derivatives we use the conventions: $x^I= (x^i, x^d$), $x^d \equiv z$,
$\partial^i=\partial_i\equiv\partial/\partial x^i$,
$\partial_z\equiv\partial/\partial z$, $\partial^\pm=\partial_\mp
\equiv \partial/\partial x^\mp$, where indices take the values
$i,j =1,\ldots, d-2$; $I,J,K,L=1,2,\ldots,d-2, d$. Vectors of the $so(d-1)$
algebra are decomposed as $X^I=(X^i,X^z)$, $X^I Y^I = X^i Y^i + X^zY^z$. %
Our fields are singlets of internal symmetries. We expect that internal symmetries can be incorporated by using the standard methods considered, e.g., in Refs.\cite{Konstein:1989ij}-\cite{Skvortsov:2020wtf}). For $n_\minrm=0$, we note the intriguing matching of field content \rf{09062025-man02-01} and the one in higher-spin theory in Refs.\cite{Vasiliev:1990en}. For the recent development in this topic, see Refs.\cite{Tatarenko:2024csa}-\cite{DeFilippi:2019jqq}. Other interesting related studies may be found in Refs.\cite{Didenko:2021vui}-\cite{Tran:2025uad}.
}
{\small
\be \label{09062025-man02-01}
\bigoplus_{n=n_{\min}}^\infty  \,\,\phi^{I_1\ldots I_n}(x,z)\,, \hspace{3cm}  \phi^{III_3\ldots I_n}(x,z) = 0\,,
\ee
}
\!\!where $n_\minrm$ depends on the type of CSF and may be read from the last column in Table I, while  fields $\phi^{I_1\ldots I_n}(x,z)$ are scalar, vector, and {\it traceless} totally symmetric tensor fields of the $so(d-1)$ algebra. To use an index free notation we introduce a unit vector $u^I$, $u^Iu^I=1$ which describes a sphere $S^{N-1}$ embedded in $\Eo^N$, $N=d-1$. The continuous-spin field $\phi(x,z,u)$ is defined as
{\small
\beq
\label{09062025-man02-05} && \phi(x,z,u) = \sum_{n=n_\minrm}^\infty \phi_n(x,z,u)\,,\quad \ \phi_n(x,z,u): = \frac{1}{ n!\sqrt{\mu_n\tau_n S_{N-1}^{\phantom{1}} } } u^{I_1} \ldots u^{I_n} \phi^{I_1\ldots I_n}(x,z)\,,
\nonumber\\
&& \hspace{1.3cm} \tau_n = \frac{ \Gamma(\frac{N}{2}) }{ 2^n\Gamma(\frac{N}{2}+n) }\,,\qquad S_{N-1} =  \frac{ 2\pi^{N/2} }{ \Gamma(\frac{N}{2}) }\,, \qquad N:= d-1\,,
\eeq
}
\!where $S_{N-1}$ stands for surface area of the sphere $S^{N-1}$ of radius 1 embedded in $\Eo^N$.

\noinbf{Light-cone gauge action and spin operators}. Light-cone gauge action for CSF takes the form%
\footnote{Light-cone gauge approach for fields in AdS was developed in Refs.\cite{Metsaev:1999ui}.  For the update presentation, see Ref.\cite{Metsaev:2019opn}.
}
{\small
\be \label{09062025-man02-10}
S  = \int dz d^dx du\, \phi^*(x,z,u) \bigl(\Box + \partial_z^2 - \frac{1}{z^2}A\bigr)\phi(x,z,u)\,, \qquad \Box = 2\partial^+\partial^- + \partial^i\partial^i \,,
\ee
}
\!where $du$ is the rotation invariant measure on the sphere $S^{N-1}$.
Operator $A$ is expressible in terms of operators $M^{IJ}$, $B^I$ which we refer to as spin operators. The operators $M^{IJ}$ obey commutators of the $so(d-1)$ algebra, while the operators  $B^I$ are AdS cousins of spin boost operators in flat space. Using the decompositions $M^{IJ}= M^{zi}, M^{ij}$, $B^I= B^z, B^i$, we note the relation for the $A$:
\be \label{09062025-man02-14}
A  = \CC_2 + 2B^z + 2M^{zi}M^{zi}+\frac{1}{2}M^{ij}M^{ij} +\frac{d^2-1}{4}\,,
\ee
where $\CC_n$ stands for an eigenvalue of the $n$th order Casimir of the $so(d,2)$ algebra. The $\CC_2$ and $\CC_4$ can be presented in terms of labels $p$ and $q$ which are complex-valued in general,%
\footnote{$p$ and $q$ are related to conformal dimension $\Delta$ and spin $s$ as $(\Delta - \frac{N+1}{2})^2=p^2$, $(s+\frac{N-1}{2})^2 = q^2$.}
{\small
\be \label{09062025-man02-17}
\CC_2 =  p^2 + q^2 - \frac{N^2+1}{2}\,,\qquad  \CC_4 = \big( p^2 - \frac{(N-1)^2}{4} \big) \big(q^2 - \frac{(N-1)^2}{4} \big)\,.
\ee
}
\!From \rf{09062025-man02-10}, \rf{09062025-man02-14}, we see that in order to fix the  action one needs to find the spin operators $M^{IJ}$, $B^I$ which should satisfy eqs.(2.13),(2.14) in Ref.\cite{Metsaev:2019opn}. Solution to the operators $M^{IJ}$ is obvious,
\be \label{09062025-man02-19}
M^{IJ} = u^I \PP^J -  u^J \PP^I\,,\qquad [\PP^I,u^J]= \theta^{IJ}\,, \qquad \theta^{IJ}:=\delta^{IJ} - u^I u^J\,,
\ee
where, $\PP^I$ stands for the derivative of the unit vector $u^I$. Finding the operators $B^I$ requires some efforts. On a space of CSF \rf{09062025-man02-05}, we find the following realization for the operators $B^I$:
{\small
\be \label{09062025-man02-21}
B^I = -(p + q)\big( \PP^I + \frac{1-N}{2} u^I\big)  - \Big( pq + \frac{(N-1)(N-3)}{4}\Big) u^I + \half \{u^I,\PP^2\}\,,
\ee
}
\!where $\{a,b\}:=ab+ba$, $\PP^2:=\PP^I\PP^I$. Remarkably, in contrast to the somewhat complicated expression for $B^I$ in the oscillator approach in Ref.\cite{Metsaev:2019opn}, operators $B^I$ \rf{09062025-man02-21} are realized as a simple differential operator. As side remarks: \abf) $B^I$ can be rewritten as {\small $ B^I = - p \PP_q^I + \half \{M^{IJ},\PP_q^J\}$, $\PP_q^I:=\PP^I + (\frac{1-N}{2}+q)u^I$}; \bbf) field $\phi_n$ \rf{09062025-man02-05} obeys the relation $(\PP^2+n(n+N-2))\phi_n=0$.

The field $\phi^*$ entering action \rf{09062025-man02-10} and an inner scalar product for field \rf{09062025-man02-05} are defined as
{\small
\be \label{09062025-man02-23}
\phi^*(x,z,u) := \sum_{n=n_\minrm}^\infty \mu_n  \phi_n^\dagger(x,z,u)\,, \qquad
(\phi,\phi): = \sum_{n=n_\minrm}^\infty \mu_n \int du\, \phi_n^\dagger(x,z,u) \phi_n(x,z,u)\,,
\ee
}
\!where $\mu_n$ are presented below, while the inner scalar product can be rewritten as  $(\phi,\phi) = \int du \phi^* \phi$.

\noinbf{Classification of classically unitary CSFs}. Using inner scalar product \rf{09062025-man02-23}, we note that, if the operators $A$, $B^I$, $M^{IJ}$ and the measure  $\mu_n$ meet the following two requirements
{\small
\be
\label{09062025-man02-27} \abf)\quad A^\dagger = A\,, \quad B^{I\dagger} = B^I\,,\quad M^{IJ\dagger} = - M^{IJ}\,, \qquad   \bbf) \quad \mu_n>0\,, \ \ \ n=n_\minrm,n_\minrm+1,\ldots,\infty, \quad
\ee
}
\!then we refer to our CSF as {\it classically unitary} CSF. The requirement $\abf)$ in \rf{09062025-man02-27} implies that light-cone gauge action \rf{09062025-man02-10} is real-valued, while the requirement $\bbf)$ in \rf{09062025-man02-27} tells us that the kinetic terms of the fields $\phi_n(x,z,u)$ should enter the action with the correct sign.

Using \rf{09062025-man02-21}, \rf{09062025-man02-27} we can find all allowed values of the labels $p$, $q$ and solution for the $\mu_n$ and hence to introduce a classification of the classically unitary CSFs. To this end, first, by requiring the $\CC_2$ and $\CC_4$ to be real-valued, we find the basic restrictions on the labels $p$, $q$,
{\small
\beq
&& \hspace{-1cm} \ibf: \ \ \Re p =0, \ \ \Re q=0; \hspace{1cm}  \iibf: \ \ p^*= q; \hspace{3cm} \iiibf: \ \ p^*= -q\,;
\nonumber\\[-11pt]
\label{09062025-man02-30} &&
\\[-9pt]
&& \hspace{-1cm} \ivbf: \ \ \Re p =0,\ \  \Im q=0; \hspace{0.8cm} \vbf: \ \ \Im p=0, \ \  \Re q=0\,; \hspace{1.1cm} \vibf: \ \ \Im p=0, \ \  \Im q=0.\qquad
\nonumber
\eeq
}
\!Second, using restrictions imposed by the classical unitarity \rf{09062025-man02-27}, we find the $\mu_n$ and additional restrictions on the labels $p$, $q$. Our results are summarized in  Table I.

\noinbf{Flat space limit}. We use the notation $B_\flatrm^I$ for the flat space cousins of the operators $B^I$ \rf{09062025-man02-21} and the notation CSF${}_\flatrm$ for CSF in the flat space. In the flat space limit, $R\sim \infty$, the asymptotic behaviour of the $B^I$ takes the form $B^I \sim R B_\flatrm^I$, where $R$ is a radius of AdS space.
For massless CSF${}_\flatrm$, $B_\flatrm^I = \kappa u^I$, $\Im\, \kappa=0$, while, for massive CSF${}_\flatrm$, {\small $B_\flatrm^I = -\irm |m| (\PP + (\frac{1-N}{2}+q)u^I)$, $\Re\, m=0$}.
In the flat space limit, we note that
\abf) for $p\sim R$, $q\sim R^0$, the series $\ibf\hbox{-}\pbf$ are reduced to massive CSF${}_\flatrm$, while for $p=-q$, $q\sim R$, to massless CSF${}_\flatrm$;
\bbf) for $p\sim R^{1/2}$, the series $\iibf\hbox{-}\pbf$, $\iiibf\hbox{-}\pbf$ are reduced to massless CSF${}_\flatrm$;
\cbf) for $p \sim R$, the series $\ivbf\hbox{-}\cbf$ and $\ivbf\hbox{-}\dbf$ are reduced to massive CSF${}_\flatrm$ of the complementary and discrete series, while, for $q\sim R$, the series $\vbf\hbox{-}\cbf$, $\vbf\hbox{-}\dbf$ are reduced to massive CSF${}_\flatrm$ of the complementary and discrete series;
\dbf) for $p \sim R^a$, $q \sim R^{1-a}$, $0< a < 1$, the series $\vibf\hbox{-}\cbf\hbox{-}1$ are reduced to massless CSF${}_\flatrm$;
\ebf) for $p \sim R^{1/2}$, the series $\vibf\hbox{-}\cbf\hbox{-}2^\pm$ are reduced to massless CSF${}_\flatrm$;
\fbf) the series  $\vibf\hbox{-}\cbf$, $\vibf\hbox{-}\dbf$, and  $\vibf\hbox{-}\dbf^*$ are not reduced to CSF${}_\flatrm$.

Now we just conjecture two of the possible definitions of masslessness for CSF in AdS.

\noinbf{Masslessness in AdS. I}. For massless CSF${}_\flatrm$, we note that \abf) the $\PP^I$-term does not contribute to $B_\flatrm^I=\kappa u^I$; \bbf) $n_\minrm=0$ in \rf{09062025-man02-01}. This motivates us to define the massless CSF in AdS by the relations $p=-q$ in \rf{09062025-man02-21} and $n_\minrm=0$ in \rf{09062025-man02-01}. From Table I, we learn then that the massless CSFs in AdS could be related to series $\ibf\hbox{-}\pbf$ and $\vibf\hbox{-}\cbf\hbox{-}2_k^-$, where, for the case of series $\ibf\hbox{-}\pbf$, we should add the additional restriction $p=-q$. Note that the $\{u^I,\PP^2\}$-term \rf{09062025-man02-21} is unavoidable in  AdS.

\noinbf{Masslessness in AdS. II}. The series $\iibf$, $\iiibf$ \rf{09062025-man02-30} are represented as the series $\iibf\hbox{-}\pbf$, $\iiibf\hbox{-}\pbf$ in Table I.  In the flat space limit, these series are reduced to massless CSF${}_\flatrm$. Therefore, in principle, the series $\iibf\hbox{-}\pbf$, $\iiibf\hbox{-}\pbf$ could also be selected as the candidates for massless CSF in AdS.

\noinbf{Comparison with oscillator formulation}. We compare the restrictions on $p,q$ in Table I and the ones in the oscillator approach in Ref.\cite{Metsaev:2019opn}. The restrictions on $p$, $q$ for the series $\ibf$, $\iibf$, $\iiibf$ in Table I and Ref.\cite{Metsaev:2019opn} coincide. For the series $\ivbf$, the oscillator formulation leads to the restrictions only on $|q|$. Therefore we compare only the restrictions on $|q|$. We then note that, for the series $\ivbf$, the restrictions on $|q|$ in Table I coincide with the ones given in (3.25), (3.33), (3.46) in Ref.\cite{Metsaev:2019opn}. If we make the replacement $q\rightarrow p$ and use (3.26), (3.34), (3.47) in Ref.\cite{Metsaev:2019opn}, then all we said for the series $\ivbf$ remains to be valid for the series $\vbf$. For the series $\vibf\hbox{-}\cbf$, $\vibf\hbox{-}\cbf\hbox{-}1$, $\vibf\hbox{-}\cbf\hbox{-}2^\pm$, the restrictions on $|p|$, $|q|$ in Table I coincide with the ones given in (3.27), (3.28), (3.29) in Ref.\cite{Metsaev:2019opn}.

We note the seeming disagreement between the restrictions on $|p|$ and $|q|$ for series $\vibf\hbox{-}\dbf\hbox{-}1$, $\vibf\hbox{-}\dbf\hbox{-}2$ in Table I and their cousins in (3.36), (3.38) in Ref.\cite{Metsaev:2019opn}.  Resolution of this puzzle is trivial. In Ref.\cite{Metsaev:2019opn}, starting with the field {\small $\phi^{0,\infty}$, $\phi^{M,K}:=\sum_{n=M}^K \phi_n$} and setting $q_n^2=x_n$,  we used the decomposition $\phi^{0,\infty} = \phi^{0,s}+\phi^{s+1,\infty}$ and  imposed the classical unitarity restrictions on the both fields - the finite component field $\phi^{0,s}$ and CSF $\phi^{s+1,\infty}$. This leads to restrictions (3.36), (3.38) in Ref.\cite{Metsaev:2019opn}. However, if we ignore the field $\phi^{0,s}$ and consider the classical unitarity restrictions only for the CSF $\phi^{s+1,\infty}$, then restrictions on $p_n^2$ in (3.38) in Ref.\cite{Metsaev:2019opn} should be replaced by the restriction $p_n^2 < x_{n+1}$. Doing so, we find the  agreement between restrictions on $|p|$, $|q|$ for series $\vibf\hbox{-}\dbf\hbox{-}1$, $\vibf\hbox{-}\dbf\hbox{-}2$ in Table I and their cousin in (3.36) in Ref.\cite{Metsaev:2019opn} and the restriction $p_n^2 < x_{n+1}$. All we said for series $\vbf\hbox{-}\dbf$ is replicated for  series $\vibf\hbox{-}\dbf^*$ in Table I and their cousins in (3.35), (3.37) in Ref.\cite{Metsaev:2019opn}.

\newpage
{\small
\noinbf {\bf Table I}. Labels $p$, $q$ and measures $\mu_n$ corresponding to the classically unitary CSF in $AdS_{d+1}$, $d>3$. Notation: $\rrr_n:= n + \frac{N-1}{2}$, $N:=d-1$. The label $n$ in $\mu_n$ takes the same values as the one in $\phi_n$ entering the expansion of $\phi$. We note the principal series $\ibf\hbox{-}\pbf$, $\iibf\hbox{-}\pbf$, $\iiibf\hbox{-}\pbf$, the complementary series $\ivbf\hbox{-}\cbf$, $\vbf\hbox{-}\cbf$, $\vibf\hbox{-}\cbf$ and the discrete series $\ivbf\hbox{-}\dbf$, $\vbf\hbox{-}\dbf$, $\vibf\hbox{-}\dbf$, $\vibf\hbox{-}\dbf^*$. We use the normalization $\mu_{n_\minrm}=1$.
\begin{center}
\begin{tabular}{|l|l|c|c|}
\hline &&&
\\[-3mm]
Series  & $p$, $q$  &  $\mu_n$ & $\phi$
\\[1mm]
\hline &&&
\\[-3mm]
\ibf\hbox{-}\pbf  & $\Re\,p=0,\ \Re\, q=0$ &  1 &
\\[1mm]
\cline{1-3} &&&
\\[-4mm]
\iibf\hbox{-}\pbf    & $p = q^*$  &   $\big|\frac{\Gamma(\rrr_n + p)\Gamma(\rrr_0 - p)}{\Gamma(\rrr_n -p)\Gamma(\rrr_0 + p)}\big|^2$ &
\\[2mm]
\cline{1-3}&&&
\\[-4mm]
\iiibf\hbox{-}\pbf & $p=-q^*$ &  1  &  $\sum\limits_{n=0}^{\infty}  \!\phi_n$
\\[-1mm]
\cline{1-3} &&&
\\[-2mm]
\ivbf\hbox{-}\cbf & $\Re\,p=0, \ \Im\, q = 0, \ |q| < r_0 $  & $\frac{ \Gamma(\rrr_n+q) \Gamma(\rrr_0-q) }{ \Gamma(\rrr_n-q)\Gamma(\rrr_0 + q)}$ &
\\[2mm]
\hline &&&
\\[-3mm]
\hspace{-0.2cm}$\qqq{ \ivbf\hbox{-}\dbf_s}{ s\in \No_0 } $ & $\Re\,p=0$, \ $\Im\, q = 0$, $q =- r_s$  & $ \frac{\Gamma(n-s)\Gamma(2s+N)}{\Gamma( n + s +N-1)} $ & $ \sum\limits_{n=s+1}^{\infty}  \!\!\!\phi_n $
\\
\hline
& \mc{3}{|c|}{}
\\[-3mm]
\vbf:  & \mc{3}{|c|}{ $ \vbf\hbox{-}\cbf = \ivbf\hbox{-}\cbf|_{ p\leftrightarrow q} $ , $ \hspace{2cm} \vbf\hbox{-}\dbf_s = \ivbf\hbox{-}\dbf_s|_{p\leftrightarrow q} $  \hspace{1.5cm}{}~}
\\[1mm]
\hline &&&
\\[-3mm]
\vibf\hbox{-}\cbf & $\Im\,p=0$, $\Im\, q = 0$, $|p| \ne |q|$,  &  &
\\
& $|p| < \rrr_0, \  |q| < \rrr_0  $ & $ \frac{\Gamma(\rrr_n+p) \Gamma(\rrr_n+q) \Gamma(\rrr_0-p) \Gamma(\rrr_0-q)}{\Gamma(\rrr_n - p) \Gamma(\rrr_n -q) \Gamma(\rrr_0+p) \Gamma(\rrr_0+q)} $ &
\\[1mm]
\cline{1-2} &&&
\\[-3mm]
\hspace{-3mm} $\qqq{ \vibf\hbox{-}\cbf\hbox{-}1_k}{k \in \No_0}$ & \hspace{-3mm} $\qqq{ \Im\,p=0, \ \Im\, q = 0, \ |p| \ne |q|,}{\rrr_k\! <\! |p| < \rrr_{k+1}, \ \rrr_k\! <\! |q| < \rrr_{k+1} } $ \hspace{-4mm} &   &
\\[4mm]
\cline{1-3} &&&
\\[-2mm]
\hspace{-0.2cm} $ \qqq{ \vibf\hbox{-}\cbf\hbox{-}2_k^+}{k \in \No_0} $ & \hspace{-3mm} $\qqq{\Im\,p=0, \ \Im\, q = 0,\ p =  q,}{|p| \ne \rrr_k} $  & $ \Big(\frac{\Gamma(\rrr_n+p)\Gamma(\rrr_0-p)}{\Gamma(\rrr_n - p)  \Gamma(\rrr_0+p)}\Big)^2 $ & $\sum\limits_{n=0}^{\infty}  \!\phi_n$
\\[1mm]
\cline{1-3} &&&
\\[-3mm]
\hspace{-0.2cm} $\qqq{ \vibf\hbox{-}\cbf\hbox{-}2_k^-}{k \in \No_0}$ & \hspace{-3mm} $\qqq{ \Im\,p=0, \ \Im\, q = 0, \ p =  - q,}{|p| \ne \rrr_k} $ & $ 1 $ &
\\[1mm]
\hline &&&
\\[-3mm]
\hspace{-0.2cm} $ \qqq{\vibf\hbox{-}\dbf\hbox{-}1_s}{s \in \No_0} $ & \hspace{-3mm} $\qqq{\Im\,p=0, \ \Im\, q = 0,\ p \ne - q,}{|p|< r_{s+1}, \  q = -r_s} $ & $ \frac{\Gamma(n-s)\Gamma(2s+N)}{\Gamma( n + s +N-1)}\frac{ \Gamma(\rrr_n + p)\Gamma(\rrr_{s+1} - p) } { \Gamma(\rrr_n - p)\Gamma(\rrr_{s+1} + p) } $ & $ \sum\limits_{n=s+1}^{\infty}  \!\!\!\phi_n $
\\[1mm]
\cline{1-3} &&&
\\[-3mm]
\hspace{-0.2cm} $ \qqq{\vibf\hbox{-}\dbf\hbox{-}2_s}{s \in \No_0} $ & \hspace{-3mm} $ \qqq{\Im\,p=0, \ \Im\, q = 0, \ p =  - q,}{q = - r_s } $  & $ 1 $ &
\\[1mm]
\hline
& \mc{3}{|c|}{}
\\[-3mm]
\vibf\hbox{-}\dbf*:  & \mc{3}{|c|}{  $ \hspace{2cm} \vibf\hbox{-}\dbf\hbox{-}1_s^* = \vibf\hbox{-}\dbf\hbox{-}1_s|_{p\leftrightarrow q} $, $ \hspace{2cm} \vibf\hbox{-}\dbf\hbox{-}2_s^* = \vibf\hbox{-}\dbf\hbox{-}2_s|_{p\leftrightarrow q} $  \hspace{1.5cm}{}~}
\\[1mm]
\hline
\end{tabular}
\end{center}
}

\noinbf{Massive spin-$s$ field, $s\in\No_0$}. Massive spin-$s$ field in $AdS_{d+1}$ is associated with series $\vibf$. For such field, we find $ \phi = \sum_{n=0}^s \phi_n$,
where $\phi_n$ and $B^I$ take the same form as in \rf{09062025-man02-05}, \rf{09062025-man02-21}. For spin-$s$ massive field, $s\in \No$, using the Pochhammer symbol $(a)_b$,  we obtain
{\small
\be  \label{09062025-man02-50}
q =  s + \frac{N-1}{2}\,, \qquad  |p| >  s + \frac{N-3}{2} \,, \quad \mu_n =  \frac{(s-n)!}{(s + N - 1 + n)_{s-n}}\frac{ (p - s- \frac{N-3}{2})_{s-n} }{ (p + \frac{N-1}{2}+n)_{s-n} } \,.
\ee
}
\!The label $p$ is related to mass as $p^2 = R^2 m^2 + (s+ \frac{N-3}{2})^2$, where the massless limit is defined as $m^2\rightarrow 0$.
For massive scalar field $\phi=\phi_0$, we get $B^I\phi=0$, $q=\frac{N-1}{2}$, $p^2= R^2m^2 + \frac{(N+1)^2}{4}$, where the massless limit is defined as $R^2 m^2\rightarrow - \frac{N(N+2)}{4}$ and we recall the BF-bound, $R^2m^2 > - \frac{(N+1)^2}{4}$.
As side remark, for spin-$s$ massless field, $s\in \No_0$, our study in Ref.\cite{Metsaev:1999ui} implies that $B^I=0$.

\vspace{-0.2cm}
\newsection{ \large Bosonic and fermionic continuous-spin fields in $AdS_4$}

\noindent {\bf Field content}. To discuss light-cone gauge bosonic and fermionic CSFs propagating in $AdS_4$  we find it convenient to use helicity fields denoted as
\be
\phi_{n+\epsilon}(x,z), \quad \epsilon = 0 \ \ \hbox{for bosons}\,,  \quad \epsilon = \half \ \ \hbox{for fermions}\,, \quad n \in \Zo\,,
\ee
where the field $\phi_{n+\epsilon}(x,z)$ has helicity equal to $n+\epsilon$. We then use an angle coordinate  $\varphi=[0,2\pi]$ which parameterizes a sphere $S^1$ and introduce the CSF denoted as $\phi(x,z,\varphi)$,
{\small
\be \label{18062025-man02-05}
\phi(x,z,\varphi) = \sum_{n} \phi_{n+\epsilon}(x,z,\varphi)  \,, \qquad  \phi_{n+\epsilon}(x,z,\varphi) := \frac{e^{\irm n\varphi}}{\sqrt{2\pi \mu_n}} \phi_{n+\epsilon}(x,z) \,,
\ee
}
\!where values of $n$ and measures $\mu_n$ depend on the type of CSF and may be found in Table II.

\noinbf{Light-cone gauge action}. Light-cone gauge action for CSF in $AdS_4$ takes the form
{\small
\beq
&& S  = \int dz d^3x d\varphi\, \phi^*(x,z,\varphi) \Big(\frac{\irm }{\partial^+}\Big)^{2\epsilon} \bigl(\Box + \partial_z^2 - \frac{1}{z^2}A\bigr)\phi(x,z,\varphi)\,,\quad  \Box = 2\partial^+\partial^- + \partial^1\partial^1 \,,\qquad
\nonumber\\
\label{18062025-man02-14} && \hspace{4cm} A  =  \CC_2 + \irm \sqrt{2}(B^\Lsm-B^\Rsm) -  2M^{\Rsm\Lsm}M^{\Rsm\Lsm} + 2\,,
\eeq
}
\!where $M^{\Rsm\Lsm}$ is a helicity operator, while $B^{\Rsm,\Lsm}$ are helicity boost operators. Eigenvalues of the 2nd and 4th order Casimirs of the $so(3,2)$ algebra $\CC_2$ and $\CC_4$ are obtained by setting $N=2$ in \rf{09062025-man02-17},
{\small
\be \label{18062025-man02-17}
\CC_2 =  p^2 + q^2 - \frac{5}{2}\,,\qquad  \CC_4 = \big( p^2 - \frac{1}{4} \big) \big(q^2 - \frac{1}{4} \big)\,.
\ee
}
\!For more details of the light-cone gauge approach in $AdS_4$, see Sec.5 in Ref.\cite{Metsaev:2022ndg}. To fix action \rf{18062025-man02-14} we have to find the operators $M^{\Rsm\Lsm}$, $B^{\Rsm,\Lsm}$ which should satisfy equations given below in \rf{21062025-man02-01}. On a space of CSF \rf{18062025-man02-05}, we find the following realization for the operators $M^{\Rsm\Lsm}$, $B^{\Rsm,\Lsm}$:
{\small
\beq
\label{18062025-man02-21} && B^\Rsm = -\frac{e^{\irm \varphi}}{\sqrt{2}} (\rh-q)(\rh-p)\,,\hspace{1cm} B^\Lsm = -\frac{e^{-\irm \varphi}}{\sqrt{2}} (\rh -1 + q)(\rh -1 + p)\,,
\nonumber\\
&& M^{\Rsm\Lsm} = -\irm \partial_\varphi + \epsilon\,, \hspace{2.7cm} \rh:= -\irm \partial_\varphi+ \half +\epsilon\,, \qquad \partial_\varphi:=\partial/\partial\varphi\,.
\eeq
}
\!The field $\phi^*$ in \rf{18062025-man02-14} and an inner scalar product for field \rf{18062025-man02-05} are defined by the relations
{\small
\be
\label{18062025-man02-23} \phi^*(x,z,\varphi) := \sum_n \mu_n  \phi_{n+\epsilon}^\dagger(x,z,\varphi)\,, \qquad (\phi,\phi): = \sum_n \mu_n \int_0^{2\pi}\!\! d\varphi\, \phi_{n+\epsilon}^\dagger(x,z,\varphi) \phi_{n+\epsilon}(x,z,\varphi)\,,
\ee
}
\!where $\mu_n$ are presented below, while the inner scalar product can be rewritten as  $(\phi,\phi) = \int du \phi^* \phi$. As side remark, for $\epsilon=0$, the operators $B^{\Rsm,\Lsm}$ \rf{18062025-man02-21} can be obtained from \rf{09062025-man02-21} by using the basis of complex coordinates for $u^I = u^\Rsm, u^\Lsm$, where
{\small
$u^\Rsm =  u e^{\irm \varphi}$, $u^\Lsm =  u e^{-\irm \varphi}$, $u:=1/\sqrt{2}$}. In such basis, we find {\small $\PP^\Rsm = \irm u e^{\irm \varphi}\partial_\varphi$, $\PP^\Lsm = - \irm u e^{-\irm \varphi}\partial_\varphi$, $\PP^2 = \partial_\varphi^2$}.

\noinbf{Classification of classically unitary CSFs}. If operators $A$, $B^{\Rsm,\Lsm}$, $M^{\Rsm\Lsm}$ and the measure $\mu_n$ meet the following two requirements
{\small
\be \label{18062025-man02-27}
\abf)\quad A^\dagger = A\,, \qquad B^{\Rsm\dagger} = B^\Lsm\,,\qquad M^{\Rsm\Lsm\dagger} = M^{\Rsm\Lsm}\,; \qquad \bbf) \quad \mu_n>0 \ \hbox{for all allowed} \ n\,,
\ee
}
\!then we refer to our CSF as {\it classically unitary CSF}. Requiring the $\CC_2$, $\CC_4$ to be real-valued we find restrictions on $p$, $q$ given in \rf{09062025-man02-30}. Using then the classical unitarity restrictions \rf{18062025-man02-27}, we find measures $\mu_n$ and additional restrictions on $p$ and $q$. Our results are given in the Table II and \rf{18062025-man02-50}.

\newpage
{\small
\noindent {\bf Table II}. Labels $p$, $q$ and measures $\mu_n$ corresponding to the classically unitary CSF in $AdS_4$. Notation: $\rrr_n:= n + \half +\epsilon$, $n\in\Zo$, where, for bosons, $\epsilon = 0 $,  while, for fermions, $\epsilon = \half $.
The label $n$ in $\mu_n$ takes the same values as the one in $\phi_{n+\epsilon}$ entering expansion of $\phi$. We note the principal series $\ibf\hbox{-}\pbf$, $\iibf\hbox{-}\pbf$, $\iiibf\hbox{-}\pbf$, the complementary series $\ivbf\hbox{-}\cbf$, $\vbf\hbox{-}\cbf$, $\vibf\hbox{-}\cbf$, the discrete series $\ivbf\hbox{-}\dbf$, $\vbf\hbox{-}\dbf$, $\vibf\hbox{-}\dbf$, and the anti-discrete series $\ivbf\hbox{-}\adbf$, $\vbf\hbox{-}\adbf$, $\vibf\hbox{-}\adbf$. Normalization of $\mu_n$: for the discrete and anti-discrete series, $\mu_{n_\minrm}=1$ and $\mu_{n_\maxrm}=1$, while, for the remaining series, $\mu_0=1$. The series not presented in Table II are given in \rf{18062025-man02-50}.

\vspace{-0.1cm}
\begin{center}
\begin{tabular}{|l|l|l|c|c|}
\hline &&&&
\\[-3mm]
Series  & $p$, $q$   & $\epsilon$ &  $\mu_n$ & $\phi$
\\[1mm]
\hline &&&&
\\[-3mm]
$\ibf\hbox{-}\pbf^\epsilon$  & $\Re\,p=0$, $\Re\, q=0$ \quad   &&  1 &
\\[0mm]
\cline{1-2} \cline{4-4} &&&&
\\[-4mm]
$\iibf\hbox{-}\pbf^\epsilon$    & $p = q^*$  & $ 0,\!\half$ & $\big|\frac{\Gamma(\rrr_n + p)\Gamma(\rrr_0 - p)}{\Gamma(\rrr_n -p)\Gamma(\rrr_0 + p)}\big|^2$ & \hspace{-0.5cm} $\sum\limits_{n=-\infty}^{\infty}\!\!\!  \phi_{n+\epsilon} \hspace{-0.2cm}$
\\[1mm]
\cline{1-2}\cline{4-4}&&&&
\\[-4mm]
$\iiibf\hbox{-}\pbf^\epsilon$ & $p=-q^*$  & &  $1$  &
\\[0mm]
\hline &&&&
\\[-3mm]
$\ivbf\hbox{-}\cbf^0$ & $\Re\,p=0, \ \Im\, q = 0, \ |q| < \half $   & $ 0 $ & $\frac{ \Gamma(n+\half+q) \Gamma(\half-q) }{ \Gamma(n+\half-q)\Gamma(\half + q)}$ & $\hspace{-0.3cm}\sum\limits_{n=-\infty}^{\infty}\!\!\!  \phi_n \hspace{-0.2cm} $
\\[0mm]
\hline &&&&
\\[-3mm]
\hspace{-0.2cm} $ \qqq{\ivbf\hbox{-}\dbf_s^\epsilon}{s\in \No_0} $  & $\Re\,p=0$, $\Im\, q = 0$, $q =- r_s$    & $0,\!\half$ & $ \frac{\Gamma(n-s)\Gamma(2s+2+2\epsilon)}{\Gamma( n + s + 1+2\epsilon)} $ & $ \hspace{-0.3cm}\sum\limits_{n=s+1}^{\infty} \!\!\! \phi_{n+\epsilon} \hspace{-0.2cm}  $
\\[1mm]
\hline &&&&
\\[-4mm]
$ \ivbf\hbox{-}\dbf^\half $ & $\Re\,p=0$, $q = 0$   & $ \half $ & 1 & $\hspace{-0.3cm}\sum\limits_{n=0}^{\infty}  \!\phi_{n+\half} \hspace{-0.2cm} $
\\[0mm]
\hline &&&&
\\[-3mm]
\hspace{-0.2cm}$ \qqq{ \ivbf\hbox{-}\adbf_s^\epsilon}{ s\in \No_0 }$  & $\Re\,p=0$, $\Im\, q = 0$, $q =- r_s$    & $0,\!\half$ & $ \frac{\Gamma(n-s)\Gamma(2s+2+2\epsilon)}{\Gamma( n + s +1+2\epsilon)} $ & $ \hspace{-0.1cm} \sum\limits_{n=-\infty}^{-s-1-2\epsilon}  \!\!\!\!\phi_{n+\epsilon} \hspace{-0.2cm}  $
\\[1mm]
\hline &&&&
\\[-3mm]
$\ivbf\hbox{-}\adbf^\half$ & $\Re\,p=0$, $q = 0$  & $ \half $ &  $1$ & $ \hspace{-0.3cm}\sum\limits_{n=-\infty}^{-1}  \!\!\!\phi_{n+\half} \hspace{-0.2cm} $
\\
\hline &&&&
\\[-3mm]
$\vibf\hbox{-}\cbf^0$ & \hspace{-0.3cm} $\qqq{\Im\,p=0, \Im\, q = 0, |p| \ne |q|}{ \small |p| < 1/2, \ |q| < 1/2} $  & 0 & $ \frac{\Gamma(n+\half+p) \Gamma(n+\half+q) \Gamma(\half-p) \Gamma(\half-q)}{\Gamma(n+\half - p) \Gamma(n+\half -q) \Gamma(\half+p) \Gamma(\half+q)} $ &  $ \hspace{-0.3cm}\sum\limits_{n=-\infty}^{\infty}  \!\!\!\phi_n \hspace{-0.2cm} $
\\[0mm]
\hline &&&&
\\[-3mm]
\hspace{-0.2cm} $ \qqq{ \vibf\hbox{-}\cbf\hbox{-}1_k^\epsilon }{ \hbox{\scriptsize $k\!+\!2\epsilon \in \No_0$} } $ \hspace{-5mm} & \hspace{-0.3cm} $\qqq{ \Im\,p=0, \ \Im\, q = 0, \ |p| \ne |q|}{\rrr_k\! < \!|p|\! <\! \rrr_{k+1}, \ \rrr_k\! <\! |q|\! <\! \rrr_{k+1}}$    & &   $ \frac{\Gamma(\rrr_n+p) \Gamma(\rrr_n+q) \Gamma(\rrr_0-p) \Gamma(\rrr_0-q)}{\Gamma(\rrr_n - p) \Gamma(\rrr_n -q) \Gamma(\rrr_0+p) \Gamma(\rrr_0+q)} $ &
\\[1mm]
\cline{1-2}\cline{4-4} &&&&
\\[-3mm]
\hspace{-0.2cm} $ \qqq{ \vibf\hbox{-}\cbf\hbox{-}2_k^{\epsilon+} }{ \hbox{\scriptsize $k\!+\!2\epsilon \in \No_0$} } $ \hspace{-5mm} & \hspace{-0.3cm} $ \qqq{\Im\,p=0, \ \Im\, q = 0, \ p =  q,}{|p| \ne \rrr_k} $ & $0,\!\half$  & $ \big(\frac{\Gamma(\rrr_n+p)\Gamma(\rrr_0-p)}{\Gamma(\rrr_n - p)  \Gamma(\rrr_0+p)}\big)^2 $ & $\hspace{-0.3cm}\sum\limits_{n=-\infty}^{\infty}  \!\!\!\phi_{n+\epsilon} \hspace{-0.2cm} $
\\[1mm]
\cline{1-2}\cline{4-4} &&&&
\\[-3mm]
\hspace{-0.2cm} $ \qqq{ \vibf\hbox{-}\cbf\hbox{-}2_k^{\epsilon-} }{ \hbox{\scriptsize $k\!+\!2\epsilon \in \No_0$} } $  \hspace{-5mm} & \hspace{-0.3cm} $ \qqq{ \Im\,p=0, \ \Im\, q = 0, \ p =  - q }{|p| \ne \rrr_k}$ && $ 1 $ &
\\[1mm]
\hline &&&&
\\[-4mm]
\hspace{-0.2cm} $\qqq{ \vibf\hbox{-}\dbf\hbox{-}1_s^\epsilon }{s \in \No_0}$  & \hspace{-0.3cm} $\qqq{ \Im\,p=0, \ \Im\, q = 0, \ p \ne - q}{|p|< r_{s+1},\ q = -r_s}$ & $0,\!\half$  & $ \hspace{-0.2cm} \frac{\Gamma(n-s)\Gamma(2s+2+2\epsilon)}{\Gamma( n + s +1+2\epsilon)} \!\frac{ \Gamma(\rrr_n + p)\Gamma(\rrr_{s+1} - p) } { \Gamma(\rrr_n - p)\Gamma(\rrr_{s+1} + p) } \hspace{-0.2cm}$ &
\\[1mm]
\cline{1-2}\cline{4-4} &&&&
\\[-6mm]
\vibf\hbox{-}\dbf\hbox{-}$2_s^\epsilon$ & $\Im\,p=0$, $\Im\, q = 0$, \ $p =  - q$, &  & $ 1 $ & $ \hspace{-0.3cm}\sum\limits_{n=s+1}^{\infty}  \!\!\!\phi_{n+\epsilon} \hspace{-0.2cm} $
\\[-3mm]
$s \in \No_0$ &    $q = - r_s$   &   & &
\\[1mm]
\hline &&&&
\\[-3mm]
$\vibf\hbox{-}\dbf^\half$ & $\Im\,p=0$, \ $|p|<1$, \ $q = 0$ & $\half$ & $ \frac{\Gamma(n+1+p)\Gamma(1-p)}{\Gamma(n+1-p)\Gamma(1+p)} $ & $\hspace{-0.3cm}\sum\limits_{n=0}^\infty \!\!\!\phi_{n+\half} \hspace{-0.2cm} $
\\[1mm]
\hline &&&&
\\[-4mm]
\vibf\hbox{-}\adbf\hbox{-}$1_s^\epsilon$ & $\Im\,p=0$, $\Im\, q = 0$, \ $p \ne - q$,&    & $\qquad \frac{\Gamma(-s-n-2\epsilon)\Gamma(2s+2+2\epsilon)}{\Gamma(  s + 1 -n )}$ &
\\[-1mm]
$s \in \No_0$ & $|p|< r_{s+1}$, \   $q = -r_s$  & $0,\!\half$ & $  \ \times\, \,\frac{ \Gamma(1-\rrr_n + p)\Gamma(\rrr_{s+1} - p) } { \Gamma(1-\rrr_n - p)\Gamma(\rrr_{s+1} + p) } $ & $ \hspace{-0.2cm}\sum\limits_{n=-\infty}^{-s-1-2\epsilon}  \!\!\!\phi_{n+\epsilon} \hspace{-0.2cm} $
\\[1mm]
\cline{1-2}\cline{4-4} &&&&
\\[-4mm]
\hspace{-0.3cm} $\qqq{\vibf\hbox{-}\adbf\hbox{-}2_s^\epsilon}{s \in \No_0} $ \hspace{-5mm} & \hspace{-0.3cm} $\qqq{ \Im\,p=0, \Im\, q = 0, \ p =  - q,}{q = - r_s}$ &  & 1 &
\\[1mm]
\hline &&&&
\\[-4mm]
\hspace{-0.1cm}$\vibf\hbox{-}\adbf^\half$ & $\Im\,p=0$, \ $|p|<1$, \ $q = 0$ &  $\half$  &  $  \frac{\Gamma(-n+p)\Gamma(1-p)}{\Gamma(-n-p)\Gamma(1+p)} $ & $ \hspace{-0.3cm}\sum\limits_{n=-\infty}^{ -1 }  \!\!\!\phi_{n+\half} \hspace{-0.2cm} $
\\[1mm]
\hline
\end{tabular}
\end{center}
}

\noinbf{Additional series for Table II}. In addition to the series in Table II, we note the series $\vbf$ obtainable from the series $\ivbf$ by using the rule $\vbf=\ivbf|_{p\leftrightarrow q}$
and the series $\vibf\hbox{-}\dbf^*$, $\vibf\hbox{-}\adbf^*$
obtainable from the corresponding series $\vibf\hbox{-}\dbf$, $\vibf\hbox{-}\adbf$ by using the just mentioned rule,
{\small
\beq
\label{18062025-man02-50} && \hspace{-1cm} \vbf\hbox{-}\cbf^0\! =\! \ivbf\hbox{-}\cbf^0|_{ p\leftrightarrow q},\,
\vbf\hbox{-}\dbf_s^\epsilon \! =\! \ivbf\hbox{-}\dbf_s^\epsilon|_{p\leftrightarrow q},\,
\vbf\hbox{-}\dbf^\half \! =\! \ivbf\hbox{-}\dbf^\half|_{p\leftrightarrow q},\,
\vbf\hbox{-}\adbf_s^\epsilon \! =\! \ivbf\hbox{-}\adbf_s^\epsilon|_{p\leftrightarrow q},\,
\vbf\hbox{-}\adbf^\half \! =\! \ivbf\hbox{-}\adbf^\half|_{p\leftrightarrow q},
\nonumber\\
&& \hspace{-1cm} \vibf\hbox{-}\dbf\hbox{-}1_s^{\epsilon*} = \vibf\hbox{-}\dbf\hbox{-}1_s^\epsilon|_{ p\leftrightarrow q},\hspace{0.8cm}
\vibf\hbox{-}\dbf\hbox{-}2_s^{\epsilon*} = \vibf\hbox{-}\dbf\hbox{-}2_s^\epsilon|_{ p\leftrightarrow q},\hspace{0.8cm}
\vibf\hbox{-}\dbf^{\half*} = \vibf\hbox{-}\dbf^\half|_{ p\leftrightarrow q},
\nonumber\\
&& \hspace{-1cm} \vibf\hbox{-}\adbf\hbox{-}1_s^{\epsilon*} = \vibf\hbox{-}\adbf\hbox{-}1_s^\epsilon|_{ p\leftrightarrow q}, \quad
\vibf\hbox{-}\adbf\hbox{-}2_s^{\epsilon*} = \vibf\hbox{-}\adbf\hbox{-}2_s^\epsilon|_{ p\leftrightarrow q},\quad
\vibf\hbox{-}\adbf^{\half*} = \vibf\hbox{-}\adbf^\half|_{ p\leftrightarrow q}.\qquad
\eeq
}
\noinbf{Massive (half)integer-spin fields}. Massive spin-$(s+\epsilon)$ fields, $s\in \No_0$, are related to series $\vibf$. For such fields, ignoring a scalar field, $s=\epsilon=0$ and using the Pochhammer symbol $(a)_b$,  we find
{\small
\beq
\label{18062025-man02-35} && \phi(x,z,\varphi) = \sum_{n=-s-2\epsilon}^s \phi_{n+\epsilon}(x,z,\varphi) \,, \qquad s + 2\epsilon\in \No\,,
\nonumber\\
&& q =  s + \half + \epsilon\,, \quad  |p| >  s - \half + \epsilon \,, \quad \mu_n = \frac{(s-n)!}{(s + 1 + n+2\epsilon)_{s-n}}\frac{ (p + \half - s -\epsilon)_{s-n} }{ (p + \half + n + \epsilon)_{s-n} } \,.\qquad
\eeq
}
\!For bosonic field, the interrelation between the label $p$ and mass is described at the end of Sec.\ref{lc-action}. For spin-$(s+\half)$ fermionic field, $s\in \No_0$, we note the relation  $|p|=m$, where the massless limit is defined as $m\rightarrow s$.

\vspace{-0.2cm}
\newsection{ \large All unitary irreps of spin algebra in $AdS_4$}

Our study above given does not provide a proof that we found all CSFs. Here, for $AdS_4$, we prove that we indeed found all CSFs. To this end we find all unitary irreps of the spin algebra. We pick up then a subset of all infinite-dimensional irreps associated with bosons and fermions and, comparing such subset with our result in Table II, we verify that we found all CSFs in $AdS_4$.

Commutators of non-linear spin algebra take the form (see Sec.5 in Ref.\cite{Metsaev:2022ndg})
{\small
\be \label{21062025-man02-01}
[M^{\Rsm\Lsm}, B^\Rsm] = B^\Rsm\,, \hspace{5mm}  [M^{\Rsm\Lsm}, B^\Lsm] = - B^\Lsm\,,
\hspace{5mm}
[B^\Rsm,B^\Lsm] = \bigl( \CC_2  -  2 M^{\Rsm\Lsm}M^{\Rsm\Lsm} + 2 \bigr)M^{\Rsm\Lsm}\,,
\ee
}
\!where $\CC_2$, $\CC_4$ are given in \rf{18062025-man02-17}. For $\CC_4$, see also (5.9) in Ref.\cite{Metsaev:2022ndg}. If $\specsf\, M^{\Rsm\Lsm}\in \Zo,\Zo+\half$ , then the operators $M^{\Rsm\Lsm}$ and $B^{\Rsm,\Lsm}$ are interpreted as helicity operator and  helicity boost operators respectively. Our study is not restricted only to the unitary irreps  with $\specsf \, M^{\Rsm\Lsm}\in \Zo,\Zo+\half$. Derivation of unitary irreps of algebra \rf{21062025-man02-01} is similar to the one for 2+1 Lorentz algebra used in Ref.\cite{barut} (see also useful Ref.\cite{Dobrev:2016gqa}). Therefore we skip details and present our result.

Let $e_n$, $n\in \Zo$, be orthonormal basis vectors of unitary irreps of algebra \rf{21062025-man02-01}, $(e_m, e_n) = \delta_{mn}$. Domains of values of $n$ for various irreps are presented in Table III. We start with the relations
\be \label{21062025-man02-05}
B^\Rsm e_n = b_n^\Rsm e_{n+1}\,, \quad B^\Lsm e_n = b_n^\Lsm e_{n-1}\,, \quad
M^{\Rsm\Lsm} e_n = (n + \varepsilon)e_n\,,\quad \phi = \sum_n \phi_{n+\varepsilon} e_n\,,
\ee
where $b_n^{\Rsm,\Lsm}$ depend on $n$, $p$, $q$, and the invariant of the irreps $\varepsilon\in \Ro$, while $\phi$ stands for a vector of the space spanned by the basis vectors $e_n$. Using \rf{21062025-man02-01}, \rf{21062025-man02-05}, and the unitarity constraints $B^{\Rsm\dagger}=B^\Lsm$, $M^{\Rsm\Lsm\dagger}= M^{\Rsm\Lsm}$, we find all unitary irreps of spin algebra \rf{21062025-man02-01}. Our results for infinite-dimensional irreps are summarized in Table III and \rf{21062025-man02-30}.

If two irreps have the same $\CC_2$, $\CC_4$, and $\specsf\, M^{\Rsm\Lsm}$ \rf{21062025-man02-05}, then such irreps are considered to be equivalent.
Irreps with $\varepsilon\in \Zo$ and $\Zo+\half$ are associated with bosons and fermions respectively. Keeping in mind the just noted equivalence,  we can verify the matching of all irreps with $\varepsilon\in \Zo, \Zo+\half$ in Table III and the ones in Table II. In other words, our classification of CSFs in Table II is complete.

\vspace{3mm}
{\small
\noindent {\bf Table III}. Labels $p$, $q$, $\varepsilon$ and expressions for $|b_n^\Rsm|^2$ corresponding to the infinite-dimensional unitary irreps of spin algebra \rf{21062025-man02-01}. Notation: $\rrr_n:= n + \half +\varepsilon$, $\rrr_n^\pm := n + \half \pm |\varepsilon|$, $ \varepsilon\in \Ro$. The $n$ in $|b_n^\Rsm|^2$ takes the same values as the one in $e_n$ entering expansion of $\phi$. We note the principal series $\ibf\hbox{-}\pbf$, $\iibf\hbox{-}\pbf$, $\iiibf\hbox{-}\pbf$, the complementary series $\ivbf\hbox{-}\cbf$, $\vbf\hbox{-}\cbf$, $\vibf\hbox{-}\cbf$, the discrete series $\ivbf\hbox{-}\dbf$, $\vbf\hbox{-}\dbf$, $\vibf\hbox{-}\dbf$, and the anti-discrete series $\ivbf\hbox{-}\adbf$, $\vbf\hbox{-}\adbf$, $\vibf\hbox{-}\adbf$. The $b_n^\Lsm$ is given by $b_n^\Lsm = b_{n-1}^{\Rsm*}$. The series not presented in Table III are given in \rf{21062025-man02-30}.

\begin{center}
\begin{tabular}{|l|l|c|c|c|}
\hline &&&&
\\[-3mm]
Series & $p$, $q$   & $\varepsilon$ &  $|b_n^\Rsm|^2 $ & $\phi$
\\[1mm]
\hline &&&&
\\[-3mm]
$\ibf\hbox{-}\pbf $  & $\Re\,p\!=\!0$, $\Re\, q\!=\!0$   &&   &
\\[0mm]
\cline{1-2}   &&&&
\\[-4mm]
$\iibf\hbox{-}\pbf $    & $p\!=\!q^*$  & $ -\half\!<\!\varepsilon\!\leq \!\half$ & $\half\big(\rrr_n^2\!-\!q^2\big) \big(\rrr_n^2\!-\!p^2\big)$ & \hspace{-0.2cm} $\sum\limits_{n=-\infty\hspace{-0.2cm}}^{\infty}\!\!  \phi_{n+\varepsilon}e_n\hspace{-0.2cm}$
\\[1mm]
\cline{1-2}&&&&
\\[-4mm]
$\iiibf\hbox{-}\pbf $ & $p\!=\!-q^*$  & &     &
\\[0mm]
\cline{1-2} &&&&
\\[-3mm]
$\ivbf\hbox{-}\cbf^\brm$ & $\Re\,p\!=\!0, \ \Im\, q\!=\!0, \ |q|\!< \!\half\!-\!|\varepsilon|$   &  & & %
\\[1mm]
\hline &&&&
\\[-3mm]
\hspace{-0.2cm} $ \ivbf\hbox{-}\dbf  $  & $\Re\,p\!=\!0$, $\Im\, q\!=\!0$, $ |q|\!=\!|\varepsilon\!-\!\half|$    &$\varepsilon > 0$ & $\half(n\!+\!1)(n\!+\!2\varepsilon)\big( \rrr_n^2\!-\!p^2\big)$ &  \hspace{-0.2cm} $\sum\limits_{n=0}^{\infty}\!\!  \phi_{n+\varepsilon}e_n\hspace{-0.2cm}$
\\[1mm]
\hline &&&&
\\[-3mm]
\hspace{-0.2cm} $ \ivbf\hbox{-}\adbf  $  & $\Re\,p\!=\!0$, $\Im\, q\!=\!0$, $ |q|\!=\!|\varepsilon\!+\!\half|$    & $\varepsilon < 0$ & $\half n(n\!+\!1\!+\!2\varepsilon)\big( \rrr_n^2\!-\!p^2\big)$ &  \hspace{-0.2cm} $\sum\limits_{n=-\infty\hspace{-0.2cm}}^{0}\!\!  \phi_{n+\varepsilon}e_n\hspace{-0.2cm}$
\\[1mm]
\hline &&&&
\\[-3mm]
$\vibf\hbox{-}\cbf^\brm$ & \hspace{-0.3cm} $\qqq{\Im\,p\!=\!0, \Im\, q\!=\!0, |p| \ne |q|,}{ \small |p| < \rrr_0^-, \ |q| < \rrr_0^-} $  & &   &
\\[0mm]
\cline{1-2} &&&&
\\[-3mm]
\hspace{-0.2cm} $ \qqq{ \vibf\hbox{-}\cbf\hbox{-}1_k^\frm}{  k \in \No_0 } \hspace{-5mm} $ \hspace{-0.5cm} & \hspace{-0.3cm} $\qqq{ \Im\,p\!=\!0, \ \Im\, q\!=\!0, \ |p| \ne |q|,}{\rrr_k^-\! < \!|p|\! <\! \rrr_k^+, \ \rrr_k^-\! <\! |q|\! <\! \rrr_k^+}$    & $ -\half\!<\!\varepsilon\!\leq \!\half$ &  $\half\big(\rrr_n^2\!-\!q^2\big) \big(\rrr_n^2\!-\!p^2\big)$   &  \hspace{-0.2cm} $\sum\limits_{n=-\infty\hspace{-0.2cm}}^{\infty}\!\!  \phi_{n+\varepsilon}e_n\hspace{-0.2cm}$
\\[4mm]
\cline{1-2} &&&&
\\[-3mm]
\hspace{-0.2cm} $ \qqq{ \vibf\hbox{-}\cbf\hbox{-}1_k^\brm}{ k\in \No_0 } \hspace{-5mm} $ \hspace{-0.5cm} & \hspace{-0.3cm} $\qqq{ \Im\,p\!=\!0, \ \Im\, q\!=\!0, \ |p| \ne |q|,}{\rrr_k^+\! < \!|p|\! <\! \rrr_{k+1}^-, \ \rrr_k^+\! <\! |q|\! <\!\rrr_{k+1}^-}$\hspace{-5mm}    & &     &
\\[4mm]
\cline{1-2} \cline{4-4} &&&&
\\[-3mm]
\hspace{-0.2cm} $ \qqq{ \vibf\hbox{-}\cbf\hbox{-}2_k}{ \hbox{$k\in \No_0$} } $ \hspace{-5mm} & \hspace{-0.3cm} $ \qqq{ \Im\,p\!=\!0, \ \Im\, q\!=\!0, \ |p| =  |q|,}{|p| \ne k + \half \pm \varepsilon} $ &&  $\half\big(\rrr_n^2\!-\!p^2\big)^2 $ &
\\[1mm]
\hline &&&&
\\[-4mm]
\hspace{-0.3cm} $\qqq{ \vibf\hbox{-}\dbf\hbox{-}0_k}{ k \in \No_0 }$ \hspace{-3mm} & \hspace{-0.3cm} $ \qqq{\Im p\!=\!0, \Im\, q\!=\!0, \ |p |\ne\! |q|,}{\rrr_k\!<\!|p|\!<\!r_{k+1}, \ |q|=\half -\varepsilon}$ &  $ \!-\frac{k+1}{2}\!<\!\varepsilon\!<\!-\frac{k}{2}$\!\! &   &
\\[3mm]
\cline{1-3} &&&$\half(n\!+\!1)(n\!+\!2\varepsilon)\big( \rrr_n^2\!-\!p^2\big)$&
\\[-3mm]
\hspace{-0.1cm} $ \vibf\hbox{-}\dbf\hbox{-}1 $  \hspace{-5mm} & $\qqq{ \Im\,p\!=\!0, \ \Im\, q\!=\!0, \ |p| \ne |q|, }{|p|< \varepsilon + \half,\ |q| =|\varepsilon-\half|}$  & $ \varepsilon>0$  & & $ \hspace{-0.3cm}\sum\limits_{n=0}^{\infty}  \!\phi_{n+\varepsilon} e_n\hspace{-0.2cm} $
\\[4mm]
\cline{1-4} &&&&
\\[-3mm]
\hspace{-0.3cm}$\qqq{\vibf\hbox{-}\dbf\hbox{-}2_k}{k\in \No_0} $ \hspace{-5mm} & $\qqq{\Im\,p\!=\!0, \ \Im\,q\!=\!0, \ |p|=|q|,}{|q|=|\varepsilon-\half|}$ & $\varepsilon \ne -\frac{k}{2} $ & $\half(n\!+\!1)^2(n\!+\!2\varepsilon)^2$ &
\\[4mm]
\hline &&&&
\\[-4mm]
\hspace{-0.4cm}  $\qqq{ \vibf\hbox{-}\adbf\hbox{-}0_k}{ k \in \No_0 }$ \hspace{-5mm}  & \hspace{-2mm}$ \qqq{\Im p\!=\!0, \Im\, q=0, \ |p |\ne |q|,}{\rrr_k\!<\!|p|\!+\!2\varepsilon\!<\!r_{k+1}, \ |q|=\!\half \!+\!\varepsilon}$ \hspace{-5mm} & $ \!\frac{k}{2}\!<\!\varepsilon\!<\!\frac{k+1}{2}$\!\! &   &
\\[3mm]
\cline{1-3} &&&$\half n(n\!+\!1\!+\!2\varepsilon)\big( \rrr_n^2\!-\!p^2\big)$&
\\[-3mm]
\hspace{-0.2cm} $ \vibf\hbox{-}\adbf\hbox{-}1 $ \hspace{-5mm} & \hspace{-0.3cm} $\qqq{ \Im\,p\!=\!0, \ \Im\, q\!=\!0, \ |p| \ne |q|, }{|p|<  \half - \varepsilon,\ |q| =|\half +\varepsilon|}$ & $ \varepsilon<0$  & & $ \hspace{-0.3cm}\sum\limits_{n=-\infty\hspace{-2mm}}^{0}  \!\!\!\phi_{n+\varepsilon} e_n\hspace{-0.2cm} $
\\[4mm]
\cline{1-4} &&&&
\\[-3mm]
\hspace{-0.3cm}$\qqq{\vibf\hbox{-}\adbf\hbox{-}2_k}{k\in \No_0} \hspace{-5mm} $\!\!\!\!\!\! & $\qqq{\Im\,p=0, \ \Im\,q=0, \ |p|=|q|,}{|q|=|\half+\varepsilon|}$ & $\varepsilon \ne \frac{k}{2} $ & $\half n^2(n\!+\!1\!+\!2\varepsilon)^2$ &
\\[4mm]
\hline
\end{tabular}
\end{center}
}

\noinbf{Additional series for Table III}. In addition to the series in Table III, we note the series $\vbf$ obtainable from the series $\ivbf$ by using the rule $\vbf=\ivbf|_{p\leftrightarrow q}$
and the series $\vibf\hbox{-}\dbf^*$, $\vibf\hbox{-}\adbf^*$
obtainable from the corresponding series $\vibf\hbox{-}\dbf$, $\vibf\hbox{-}\adbf$ by using the just mentioned rule,
{\small
\beq
\label{21062025-man02-30} && \hspace{-10mm} \vbf\hbox{-}\cbf^\brm = \ivbf\hbox{-}\cbf^\brm|_{ p\leftrightarrow q},  \qquad
\vbf\hbox{-}\dbf  = \ivbf\hbox{-}\dbf|_{p\leftrightarrow q}, \qquad
\vbf\hbox{-}\adbf  = \ivbf\hbox{-}\adbf|_{p\leftrightarrow q}, \qquad
\nonumber\\
&& \hspace{-10mm} \vibf\hbox{-}\dbf\hbox{-}0_k^{\epsilon*}\!\! =\! \vibf\hbox{-}\dbf\hbox{-}0_k^\epsilon|_{ p\leftrightarrow q},
\
\vibf\hbox{-}\dbf\hbox{-}\!1^* \!\!=\! \vibf\hbox{-}\dbf\hbox{-}1|_{ p\leftrightarrow q},%
\
\vibf\hbox{-}\adbf\hbox{-}0_k^{\epsilon*} \!\!=\! \vibf\hbox{-}\adbf\hbox{-}0_k^\epsilon|_{ p\leftrightarrow q},
\
\vibf\hbox{-}\adbf\hbox{-}1^* \!\!=\! \vibf\hbox{-}\adbf\hbox{-}1|_{ p\leftrightarrow q}.\hspace{0.8cm}
\eeq
}

\vspace{-2mm}
\noinbf{Massive (half)integer-spin unitary irreps}. Massive spin-$s$ unitary irreps, $s\in \No_0,\No_0+\half$, are associated with series $\vibf$ having $\varepsilon=-s$. For such irreps, ignoring a scalar irrep, $s=0$, we find
{\small
\be
|q| = s+\half\,, \quad |p| > s-\half\,, \quad |b_n^\Rsm|^2 = \half (n+1)(2s-n) \big( p^2 - (n - s +\half)^2 \big)\,,\quad b_n^\Lsm = b_{n-1}^{\Rsm*}\,,
\ee
}
\!$n=0,1,\ldots,2s$. The irreps with $s \in \No_0$ and $s \in \No_0+\half$ are associated with spin-$s$ bosonic and spin-$s$ fermionic fields respectively. For scalar irrep, $s=0$, we note the relations   $b_n^{\Rsm,\Lsm}=0$, $|q|=\half$, $|p|^2=\frac{9}{4}+R^2m^2$ and the BF-bound $R^2m^2 > - \frac{9}{4}$. Field in $AdS_4$ associated with spin-$s$ massive irrep is given by
{\small
\be \label{07062025-01}
\phi(x,z) = \sum_{n=0}^{2s} \phi_{n-s}(x,z) e_n\,.
\ee
}
As side remark, for massless spin-$s$ irreps, $s\in \No_0$, $\No_0+\half$, we note the relations $B^\Rsm=0$, $B^\Lsm=0$.

{\bf Conclusions}. In this paper, we developed the light-cone gauge vector superspace formulation of CSF and applied such formulation for finding the spin operators and classification of classically unitary CSF in $AdS_{d+1}$, $d\geq 3$.
In view of a simple form of the spin operators obtained we expect that our approach could be helpful for a study of interacting CSF in AdS.
To this end the method for a study of interacting integer-spin field in AdS developed in Ref.\cite{Metsaev:2018xip} upon a suitable adaptation to CSF might be helpful. A light-cone gauge approach has actively been used for the study of AdS/CFT correspondence (see, e.g., Refs.\cite{Skvortsov:2018uru,deMelloKoch:2024juz}). We believe that our results in this paper might be helpful for the study of AdS/CFT of CSF along the line of the conjecture in Ref.\cite{Metsaev:2019opn}.

\vspace{-3mm}

\end{document}